\documentclass{article}
\usepackage{emulateapj,graphics,epsfig,onecolfloat}

\begin{document}
\twocolumn
[
\title{Spectroscopic detection of Type Ia Supernovae in the Sloan Digital Sky Survey}

\author{Darren S. Madgwick$^{1,2}$,
Paul C. Hewett$^{3}$, Daniel J. Mortlock$^{3}$ \& Lifan Wang$^{1}$}
\affil{$^1$ Lawrence Berkeley National Laboratory, MS50R-5032, Berkeley, CA 94720}
\affil{$^2$ Hubble Fellow}
\affil{$^3$Institute of Astronomy, Madingley Road, Cambridge 
CB3 0HA, U.K.}

\begin{abstract}

We present the results of a novel new search of the first data-release from the Sloan
Digital Sky Survey (SDSS-DR1) for the spectra of supernovae.  
The use of large
spectroscopic galaxy surveys offers the prospect of obtaining improved estimates
of the local supernova rate, with the added benefit of a very different selection
function to that of conventional photometric surveys.  In this {\em Letter} we
present an overview of our search methodology and the details of 19 Type Ia
supernovae found in SDSS--DR1.  The supernovae sample is used to make a
preliminary estimate, $\Gamma_{\rm Ia} = 0.4\pm0.2 h^2$ SNu, of the cosmological SNe
rate.
\end{abstract}

]

\section{Introduction}

Supernovae (SNe) are generally discovered through difference imaging techniques
(see e.g.\ Perlmutter et al.\ 1995; Schmidt et al.\ 1998).  However, it is also
possible to detect them spectroscopically -- by identifying the broad peaks and
troughs that typically distinguish a SN spectrum from that of its host--galaxy.

The spectroscopic approach to the detection of SNe has a number of advantages;
namely that a detection can be made with only {\em one} observation, and that
the type is readily identifiable without additional follow--up observations.
Despite these advantages it is clear that a dedicated spectroscopic SN survey is
completely impractical -- the rate of detection relative to telescope time is
far too low.  Fortunately, the data required for such a survey are already
available from several large galaxy redshift surveys now approaching completion.
In a forthcoming paper, Mortlock, Madgwick \& Hewett~\cite{Mor03} calculate that the
$10^6$ galaxy spectra to be obtained by the Sloan Digital Sky Survey (SDSS, York
et al.~\cite{Yor00}) should yield $\sim 200$ SNe detections.

One immediate application for such a sample is that the cosmological SNe rate, 
$\Gamma$, can be constrained to $< 10$ per cent, a considerable improvement 
on existing single survey measurements of $\Gamma$ (e.g. Pain et al.~\cite{Pai96}, 
\cite{Pai02}; Hardin et al.~\cite{Har00}). Moreover, using a galaxy redshift survey ensures 
that the host galaxy sample is extremely well defined, including redshifts,
broad--band colours, spectroscopic and morphological types. 

The SNe presented in this {\em Letter} have typically been observed $\sim$2--3 years previously, 
and so follow-up observations are not possible.  However, it is important to stress that 
if a search for SNe is carried out as soon as the spectroscopic data
in a given redshift survey is reduced there is potentially a significant amount 
 of time in which more detailed follow-up observations can be carried out.

This {\em Letter} describes the first results of an analysis of the $\sim 10^5$
galaxy spectra in the SDSS Data Release 1 (SDSS-DR1 Abazajian et
al.~\cite{Aba03}).  The SNe detection procedures are described in \S 2, with the
efficiency of the technique assessed using simulations in \S 3.  The actual SNe
discoveries are detailed in \S 4 before an initial estimate of the cosmological
SN rate is presented in \S 5.  The main results and future possibilities are
summarised in \S 6.

\section{Method}

The spectroscopic detection of SNe relies upon the ability to
distinguish the broad spectral features typical of these objects in a
composite host--galaxy+SN spectrum where the galaxy spectrum will usually
dominate. Two key tasks need to be addressed in order to achieve this goal:
\begin{enumerate}

\item The host--galaxy spectrum needs to be `subtracted' from the observed
spectrum, leaving a `residual' spectrum.  The residual spectrum should be
dominated either, by noise (if the original spectrum was simply that of a
galaxy), or, by SNe spectral features if these are present.

\item The `residual' spectrum then needs to be searched for the presence of
characteristically broad SNe features, which must also be distinguished
from any other artifacts that have not been removed in the previous step.
\end{enumerate}
We outline here the steps we have adopted to carry out these tasks.

\subsection{Galaxy subtraction}

Although galaxy spectra appear superficially to exhibit a wide variety of forms,
they can be very successfully parameterised with only a handful of components.
In particular, the use of a principal component analysis (PCA) has proved to be
well adapted to this task (see e.g.  Connolly et al.~\cite{Con95}; Folkes,
Lahav \& Maddox~\cite{Fol96}; Madgwick et al.~\cite{Mad02}).

The key result of the PCA is the construction of an orthogonal set of
`eigenspectra' -- each of which successively incorporates the maximum amount of
information about the observed galaxy spectral distribution.  Previous analyses
of samples of galaxy spectra have shown that as few as 8 such eigenspectra can
be used to accurately reconstruct most galaxy spectra (Folkes, Lahav \&
Maddox~\cite{Fol96}).

In our analysis we perform PCA on a representative sample of 1000 galaxy spectra
drawn from the SDSS, to construct a set of 20 eigenspectra.  The eigenspectra
provide an optimal basis for reconstructing, and then subtracting, each
host--galaxy spectrum.  The extended wavelength range, excellent resolution and
spectrophotometric qualities of the SDSS spectra mean that all 20 components
contain information of significance for the spectroscopic reconstructions.
However, our technique is not sensitive to the exact number of components
employed and the use of only 10 eigenspectra yielded essentially identical
results.

The detailed procedure is as follows:

\begin{enumerate}
\item The basis of 20 eigenspectra was constructed using PCA on 1000
galaxy spectra, chosen not to contain a strong signature of SNe
spectra or any other spectroscopic artifact or anomaly.
\item Each observed spectrum in the SDSS--DR1 was then projected onto
this basis.  These projections quantify the relative importance of different 
components in the spectra of the galaxies and the observed spectra can
be reconstructed very accurately from the 20 eigenspectra. Unusual
spectroscopic features (such as those of SNe) do not contribute to the
form of the eigenspectra and are therefore not included in the
reconstructed galaxy spectra.
\item Each galaxy spectrum was then reconstructed from the eigenspectra.
Subtracting the reconstructed spectrum from the original galaxy spectrum
produces a `residual' spectrum.  It is this residual spectrum that will be
searched for the SNe features.
\end{enumerate}

\subsection{SNe matching}

The residual spectrum left after subtracting the reconstructed host--galaxy
spectrum is almost always dominated by noise, validating the effectiveness of
the PCA--reconstructions.  However, spectroscopic artifacts were present for a
small fraction of the objects due to poor sky subtraction or residual nebular
emission lines, the latter resulting from particularly unusual emission line
ratios.  To reduce the impact of these relatively high--frequency residuals, a
low--pass filtering of the residual spectra was applied before attempting to
identify SN features.  The filtering was based upon the {\em $\acute{a}$ trous}
wavelet transform (Starck, Siebenmorgen \& Gredel~\cite{Sta97} and references
therein), as described in Madgwick et al.~\cite{Mad02b}.  This additional
filtering leaves a low--pass filtered residual spectrum, $f_\lambda$, allowing
us to isolate just the broadest spectral features without contamination from
narrow peaks and troughs present in the original observed spectrum.

In order to quantify the significance of a SN detection we use two parameters --
the correlation coefficient (here denoted by CC) and the RMS signal.  The first
of these parameters, CC, is determined simply by calculating the covariance
between each low--pass filtered residual spectrum, $f_i$ ($i=0,..\; N_{\rm
gal}$) and a sequence of different SN templates, $t_j$ ($j=0,..\; N_{\rm
temp}$)\footnote{We make use of the observed spectra from SN1998aq at $-$9, 0,
7, 19, 31 and 55 days (Branch et al.~\cite{Bran03}), SN1998bu at $-$4, 8, 13, 28 days (Jha et al.~\cite{Jha99}) and SN1999ee at
$-$8, $-$1, 12, 20, 33 days (Hamuy et al.~\cite{Ham02}), 
as these cover all available spectral features
uniformly over a wide range of times.},

\begin{equation}
\rm{CC_{ij}} = \frac{{\rm{Cov}}_{f_it_j}}{\sigma_i \sigma_{j}}\;.
\end{equation}
For each galaxy $i$, we set ${\rm C}_i = \rm{max}({\rm C}_{ij})$, so
that only the `best--fitting' SN template is used to determine the
likelihood of a given detection. The resulting value of CC quantifies
the similarity between the shape of the residual spectrum and the shape of
the best--fitting SN template.

The RMS signal is also calculated from the low--pass filtered residual
spectrum.  This parameter provides a measure of the amplitude of the
signal present in the residual spectrum.

\begin{equation}
{\rm RMS} = \sqrt{ \left( \frac{\sum_{\lambda} f_{\lambda}}{N_{\lambda}}\right)^2 } \;.
\end{equation}

\section{Simulations}

There are several criteria which determine whether a SN will be detectable in a
given spectrum.  One of the most important of these is the relative brightness
of the SN compared with the host-galaxy -- which will determine the prominence
of the SN features in the observed composite host--galaxy+SN spectrum.

In order to test how important these effects are we simulated several thousand
composite `host--galaxy+SN' spectra, by randomly choosing objects from the
SDSS--DR1 and adding our SNe templates.  The flux in the galaxy spectra and SN
templates was calculated over the wavelength range $5000-6000$\AA, i.e.
approximating the $V$-band, and SNe templates were added with flux ratios
(SN/galaxy) in the range 0.75-0.1.  For each simulated spectrum we then ran our
detection algorithm and calculated both the CC and RMS values.  
%
It was determined from these simulations that our algorithms work very well at identifying a SN
component in the spectra if the SN contributes $\ge 10\%$ of the $V$--band flux
recorded in the SDSS fibre--spectrum (more details will be presented in 
Mortlock, Madgwick \& Hewett~2003).


An empirical cut in CC and RMS allows
the detection of virtually all SNe present down to a relative flux ratio of 0.1
(corresponding to a magnitude difference $\Delta m\simeq 2$ between the
host-galaxy and SN).

\section{Results}

We have calculated the CC and RMS parameters for all 115,977 objects
in SDSS--DR1 with $z\le 0.25$, $z_{\rm confidence}\ge 0.35$ and 
a spectroscopic classification of `galaxy'.

Some 1500 SDSS--DR1 spectra satisfied our selection criteria,
representing $\sim1$\% of this spectroscopic galaxy sample.  However, although
we inspected visually all of these spectra, high--confidence detections of SN
were only made for objects which had $\rm{CC} \ge 0.6$ -- corresponding to only
the 230 most significant detections. The detection algorithm has proved
highly successful in identifying spectra with SNe present at high confidence.

In total, 19 Type Ia SNe were identified with high confidence from our search.
Another 15 possible detections were made, although further work is required both
to quantify the true fraction of SNe in this sub--sample and to optimise the
detection efficiency for the presence of SNe at low flux ratios.  Properties of
the high--confidence SNe detections are given in Table~\ref{tab:res}.

\begin{table*}
\begin{center}
\caption{
Type Ia SNe discovered in SDSS--DR1.  In each case  the SNe type, magnitude and ages
have been derived using a wide variety of SNe templates (including 35 
Type Ib/c and  41 Type Ia spectra).  
The quoted ages are the $\pm 1\sigma$ range about the expected
value.  Also given are the flux ratios between the galaxy and SN components of
each fibre--spectrum -- as derived from the same template library.
\label{tab:res}}
\begin{tabular}{@{}cccccccc@{}}
\hline
Name & R.A. & Dec & $V_{\rm SN}$ mag & Flux ratio & z & 
Date observed & SN age (days) \\
\hline
2000fz & 01 18 35.83 & +14 41 00.5 & 17.8  & 0.40 &  0.054 & 2000-12-15 & 6.3 -- 6.8 \\
2000ga & 02 29 03.73 & $-$08 24 13.7 & 20.0  & 0.14 &  0.141 & 2000-12-24 & 6.0 -- 7.9 \\
2001kl &  03 08 51.44 & $-$01 10 24.1 & 20.4  &  0.06 & 0.126  & 2001-01-22 & $\sim20$ \\
2001ks &  07 56 46.50 & +36 59 17.0 & 18.7  & 0.12 &  0.078 & 2001-04-18 & -5.0 -- -3.0 \\
2000fy &  08 03 12.61 & +47 36 49.7 & 19.5  & 0.19 &  0.117 & 2000-12-06 & 2.5 -- 3.5 \\
2001kn &  08 32 09.85 & +47 17 27.6 &  20.0   &  0.17    &   0.134  & 2001-03-13 &  -5.0 -- 3.0 \\
2001km &  09 11 38.38 & $-$00 42 54.0  & 18.4  &  0.56 & 0.070 & 2001-02-15 & 6.0 -- 6.3 \\
2001kj &  09 22 29.15 & +57 54 29.3 & 18.1  &  0.30 & 0.063 & 2001-01-02 & 10 -- 11 \\
2001kq &  09 33 34.31 & +55 10 26.0 & 18.5  &  0.08 & 0.074 & 2001-03-23 & -4.9 -- -3.1 \\
2001kp &   09 51 53.08 & +01 06 05.8 & 18.2  & 0.34 &  0.063 & 2001-03-21 & -3.0 -- -2.0 \\
2001kr &  09 59 15.75 & +00 58 02.4 & 19.1  & 0.43 &  0.088 & 2001-03-26 & 11-- 12 \\
2000fx &  10 18 00.47 &  $-$00 01 58.0 & 18.7  &  0.26 &   0.065 & 2000-12-01 & 16 -- 18 \\
2001ki & 10 48 58.40 & $-$00 26 45.2 & 19.3 & 0.14 &  0.106 & 2001-01-01 & 6.0 -- 7.3 \\
2001kk & 12 47 33.41 & +00 05 57.1 & 18.8  & 0.15 &  0.086 & 2001-01-19 & 5.7 -- 7.8 \\
2001ku &  13 39 44.77 & +62 23 37.0  & 19.9  &  0.15 & 0.136 & 2001-06-19 & 3.3 -- 4.0 \\
2001ko &  14 10 58.32 & +64 50 50.9   & 20.0  & 0.15 &  0.141 & 2001-03-16 & 7.6 -- 8.4 \\
2000fw &  14 30 14.07 &  +00 30 35.4  & 19.1  &  0.28 & 0.096 & 2000-03-10 & -4.6 -- -3.3 \\
2001kt &  16 17 13.40 & +48 28 27.7 & 19.6 & 0.30 & 0.104 & 2001-05-25 & 14 -- 17 \\
2000gb & 17 32 28.54  & +56 04 25.5 & 20.4 & 0.22 & 0.123 & 2000-10-01 & 16 -- 20 \\
\hline
\end{tabular}
\end{center}

\end{table*}

\begin{figure*}
\begin{center}
    \epsfig{file=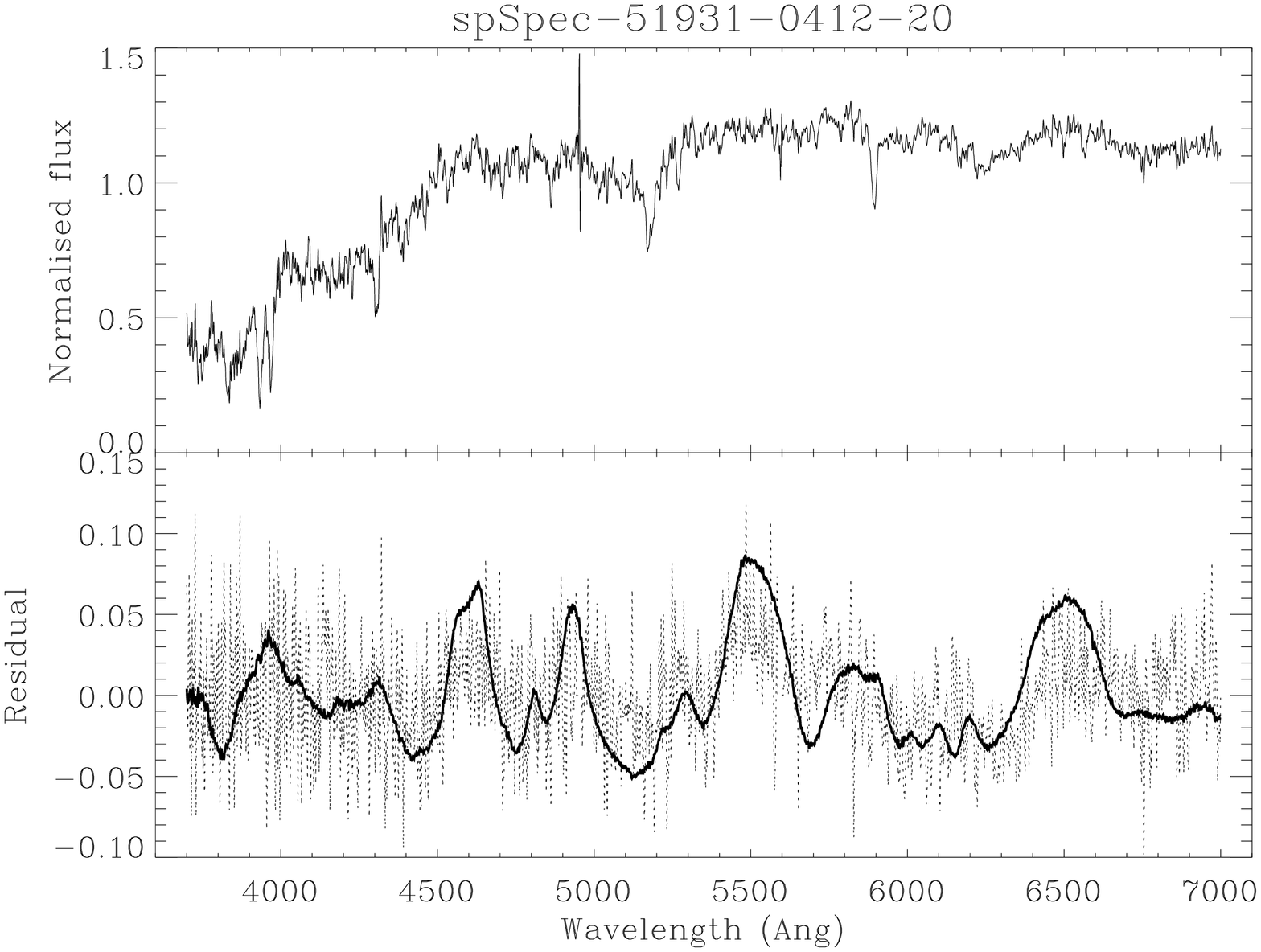,angle=90,width=5.5cm}
    \epsfig{file=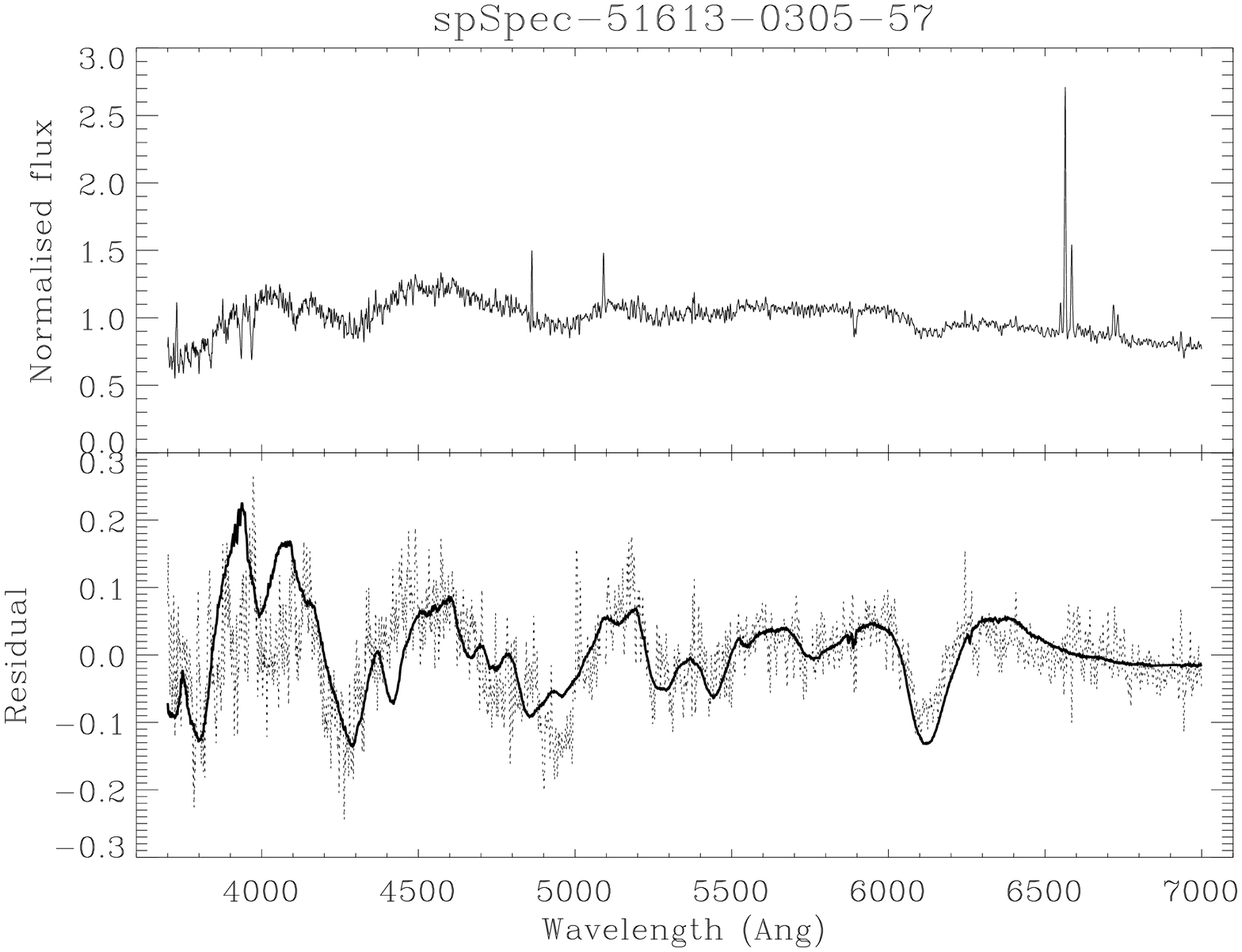,angle=90,width=5.5cm}
    \epsfig{file=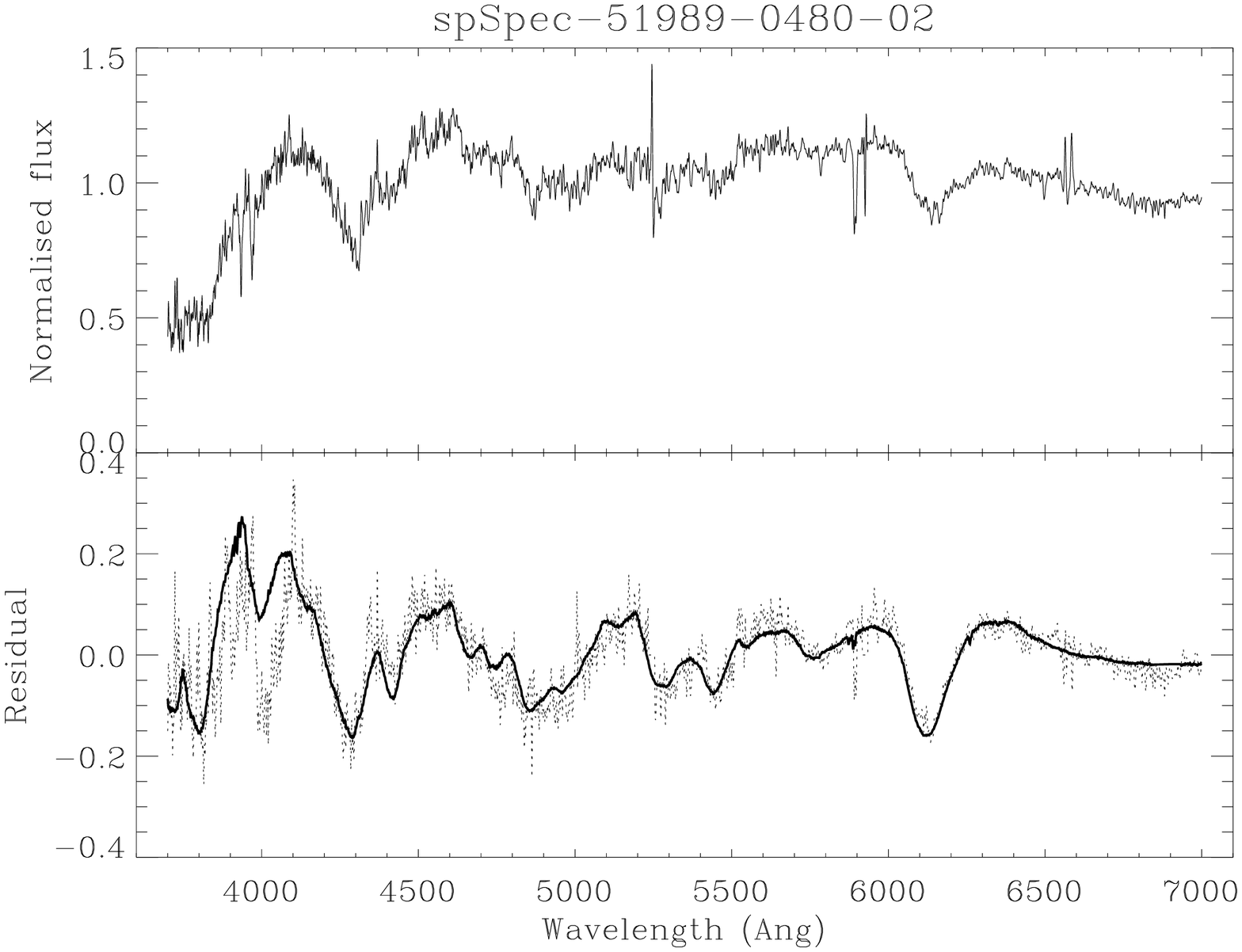,angle=90,width=5.5cm}
    \epsfig{file=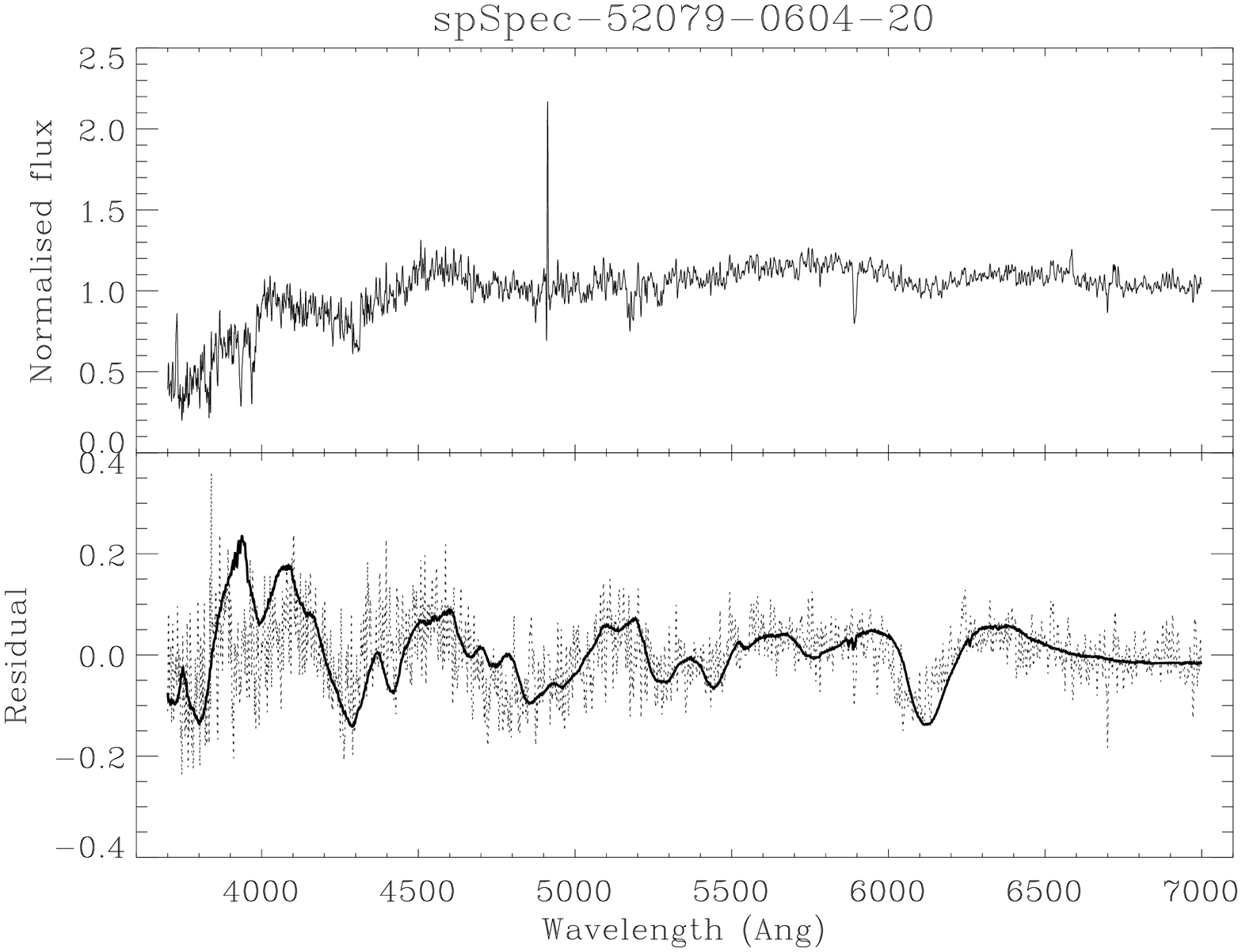,angle=90,width=5.5cm}
    \epsfig{file=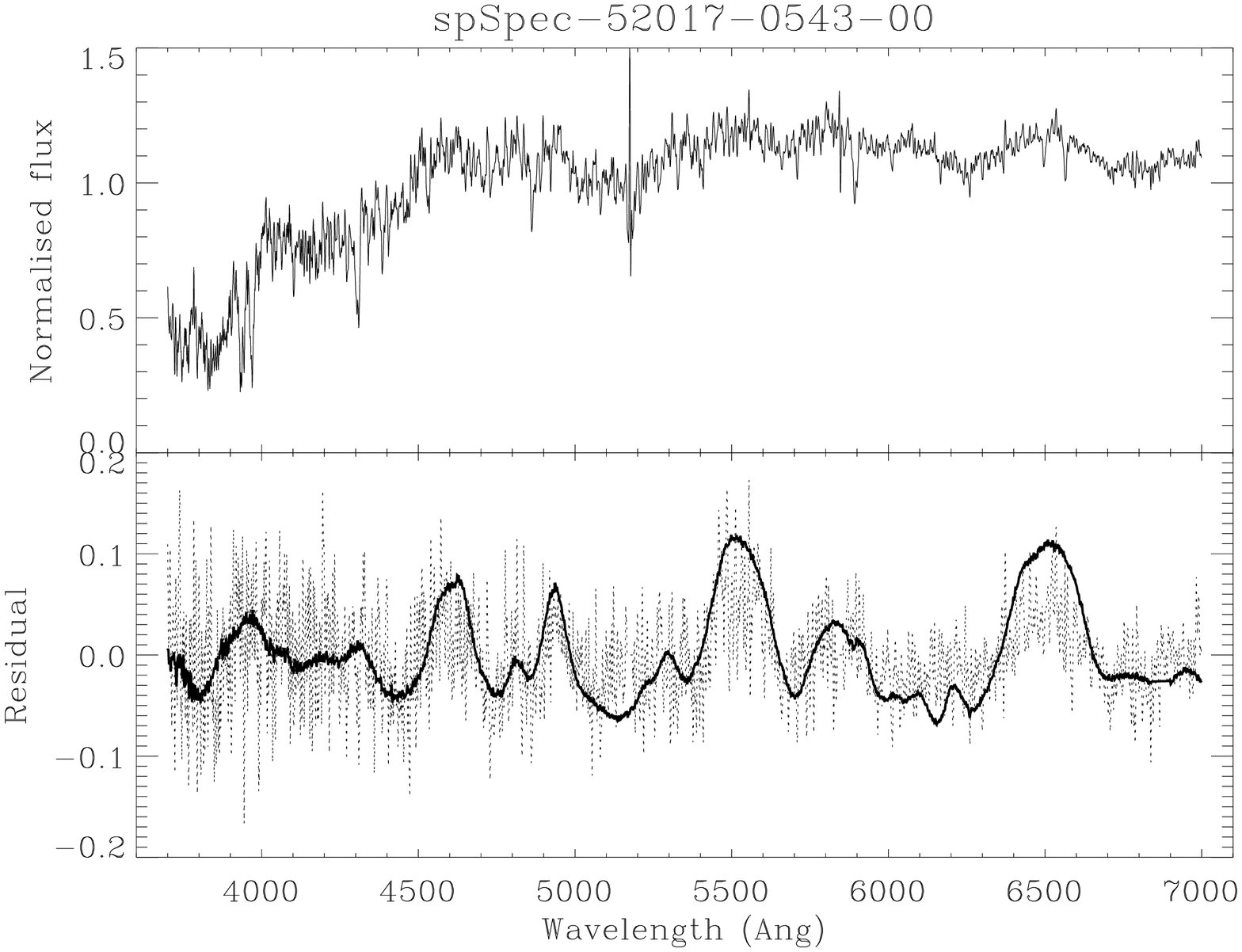,angle=90,width=5.5cm}
    \epsfig{file=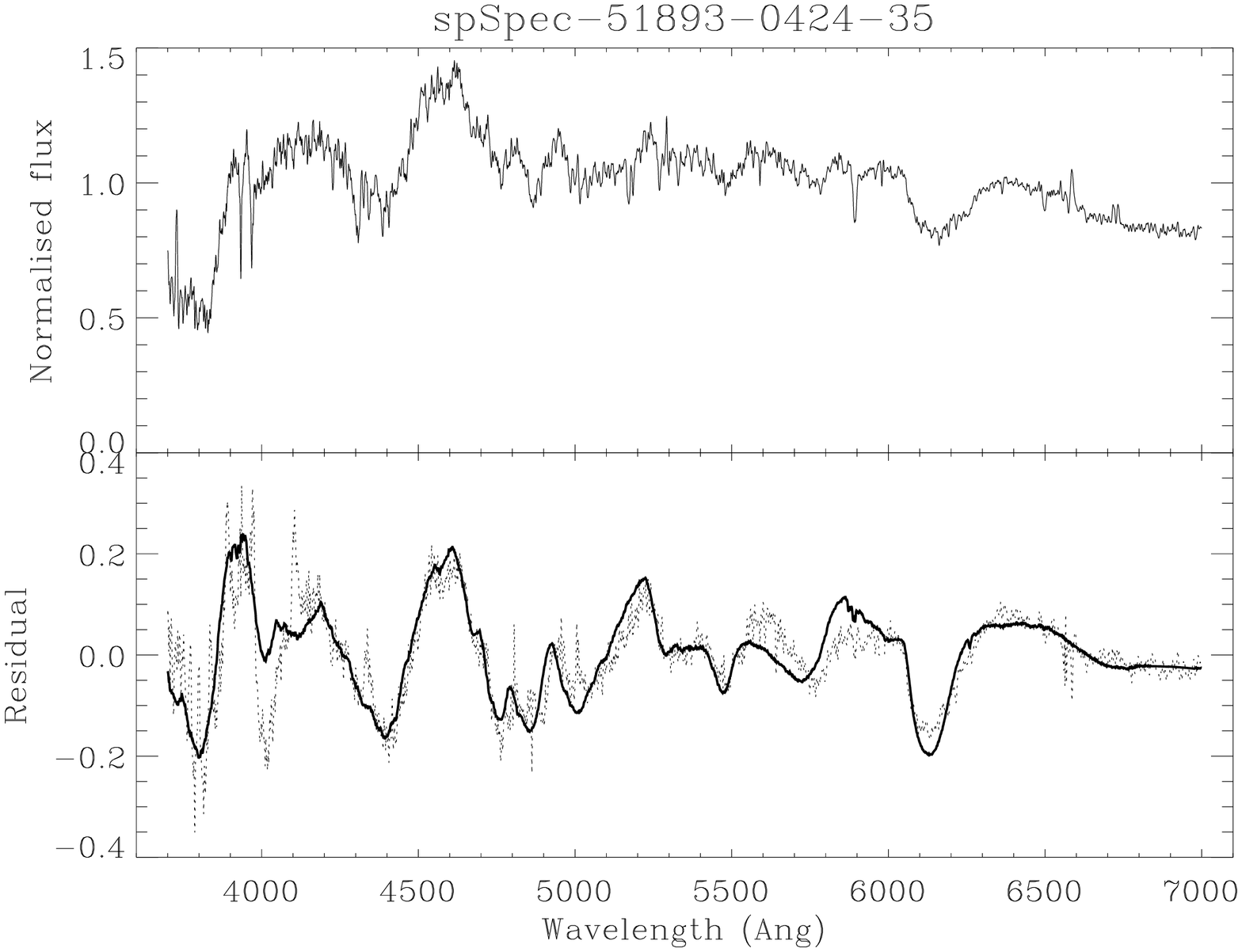,angle=90,width=5.5cm}

\caption{Example spectra of the SNe identified in SDSS--DR1 are
shown in each panel.  In addition, the lower plot in each panel shows
the `residual' spectrum that is left after subtracting the host--galaxy
from the observed spectrum.  The best--fitting template SN spectrum is
shown overplotted on this residual spectrum.
\label{fig:spec}}
\end{center}
\end{figure*}

\section{The cosmological SNe Rate}

A crude estimate of the number of SNe expected in the survey, $\langle N_{\rm
SN} \rangle$, is given by the product of four quantities:  the number of
galaxies in the survey, $N_{\rm g}$; the average luminosity of a survey galaxy,
$\langle L_{\rm g} \rangle$; the typical control time for which a SN is
sufficiently bright to be detectable, $\langle t_{\rm SN} \rangle$; and, of
course, the SN rate itself (which is conventionally measured in supernova units,
SNu\footnote{1 SNu is defined as the number of SNe per century per $10^{10}$
Solar luminosities in the $B$-band (i.e.  1 SNu = $10^{-12}$
yr$^{-1}L\odot^{-1}$)}).  Taking typical values for these quantities yields

\begin{equation}
\label{eqn:n_sn}
\langle N_{\rm SN} \rangle
  \simeq
  20
  \frac{N_{\rm g}}{10^5}
  \frac{\langle L_{\rm g} \rangle}{L_*}
  \frac{\langle t_{\rm SN} \rangle}{10\; {\rm days}}
  \frac{\Gamma}{1\; \rm SNu}
  ,
\end{equation}

where $L_* \simeq 10^{10} L_\odot$ (Madgwick et al.~\cite{Mad02}) is the
fiducial $B$-band luminosity of a field galaxy.  The most difficult parameter to
estimate is $\langle t_{\rm SN} \rangle$, which is determined from a combination
of the SNe light--curve shapes, the positional distribution of the SNe in their
host--galaxies, the size of the apertures used to obtain the spectra and the
detection algorithm used.  A value of $\langle t_{\rm SN} \rangle
\simeq 20$ days is consistent both with the best fit ages of the SNe listed in
Table~\ref{tab:res} and the more theoretical approach taken by Mortlock, 
Madgwick \& Hewett~\cite{Mor03}, 
but is probably uncertain by up to 50\% in this 
preliminary analysis.

The estimate of the SN rate from the
discovery of 19 type Ia SNe in the SDSS--DR1 galaxy spectra is thus
\begin{equation}
\Gamma_{\rm Ia} = 0.4 \pm 0.2 h^2 \; {\rm SNu},
\end{equation}
where both the Poisson uncertainty in the sample size and our estimated
uncertainty in the control time, $\langle t_{\rm SN}\rangle$, are included.  $h$
is the value of the Hubble Constant in units of $100 \, {\rm km s^{-1}
Mpc^{-1}}$.  Our estimate is consistent with previous measurements at high redshifts:
$\Gamma_{\rm Ia}(z=0.4) = 0.8 \pm 0.5 h^2$ SNu (Pain et al.~\cite{Pai96}),
$\Gamma_{\rm Ia}(z=0.5) \simeq 0.58 \pm 0.2 h^2$ SNu (Pain et al.~\cite{Pai02}), and is
very similar to that previously determined at low-$z$:
$\Gamma_{\rm Ia}(z=0.1) = 0.4 \pm 0.3 h^2$ SNu (Hardin et al.~\cite{Har00}).

\section{Conclusions}

A novel spectroscopic analysis of the $\sim 10^5$ galaxy spectra in the SDSS-DR1
has resulted in the definite identification of 19 Type Ia SNe, along with a similar
number of less certain detections.  The resultant estimate of the local Type Ia
SN rate, $\Gamma_{\rm Ia} \simeq 0.4 \pm 0.2 h^2\,$SNu, is consistent with that
obtained by more conventional imaging methods, although the uncertainties in all
cases are dominated by the small number of SNe.

The future for spectroscopic SNe searches is promising, as this methodology
 can be applied to any redshift survey for which a large sample of galaxy spectra exist,
 opening up new scientific applications for present and future surveys.  
A more complete
analysis of the SDSS--DR1 spectra will include a more sophisticated estimate of
the Cosmological SNe rate (Mortlock, Madgwick \& Hewett~\cite{Mor03}), along with a
number of consistency checks in which the magnitudes, redshifts and epochs
(relative to maximum light) of the detected SNe are compared to the predicted
distributions.

Looking ahead, the ease with which these Type Ia SNe have been discovered in the
SDSS--DR1 spectra implies that the full survey of $\sim 10^6$ galaxies should
yield at least $\sim 200$ SNe of all types.  Such a sample should provide an
estimate of $\Gamma$ which is largely free of systematic errors and subject to
only a small statistical uncertainty.  Furthermore, the spectroscopic selection
applied to the SDSS sample will allow the frequency of SNe events in terms of
the properties of the underlying galaxy population to be determined to
unprecedented accuracy (see e.g.  Cappellaro et al.~\cite{Cap97}; Cappellaro,
Evans \& Turatto~\cite{Cap99}).

A final  important point to make is that all of these SNe have been discovered several years
after their occurrence, and as such follow-up observations are simply not possible.  However,
by incorporating our search methodology into the SDSS (and other survey's) data reduction
pipelines these objects can be discovered essentially immediately.  
It is hoped that the presentation of the results  
in this {\em Letter} will result in this occurring, so that this interesting aspect of galaxy redshift 
surveys no longer goes neglected.

\acknowledgements

DSM would like to thank Saul Perlmutter, Alex Kim and Eric Linder for many
useful discussions, and the referee David Branch for helpful
suggestions.  Support for this work was provided by NASA through Hubble
Fellowship grant \#HST-HF-01163.01-A  
awarded by the Space Telescope Science Institute,
which is operated by the Association of Universities for Research in
Astronomy, 
Inc., for NASA, under contract NAS 5-26555.  DJM was supported by PPARC.

Funding for the Sloan Digital Sky Survey (SDSS) has been provided by the Alfred
P.  Sloan Foundation, the Participating Institutions, the National Aeronautics
and Space Administration, the National Science Foundation, the U.S.  Department
of Energy, the Japanese Monbukagakusho, and the Max Planck Society.  The SDSS
Web site is http://www.sdss.org/.  The SDSS is managed by the Astrophysical
Research Consortium (ARC) for the Participating Institutions.

\end{document}